\documentclass[twocolumn]{miel}
\usepackage[dvips]{graphicx}
\usepackage[dvips]{epsfig}
\usepackage[dvips]{color}

\def\##1{{\bf #1}}
\def\=#1{\underline{\underline #1}}
\def\4#1{\underline{\underline{\underline{\underline #1}}}}

\def\.{\mbox{ \tiny{$^\bullet$} }}

\def\le{\left(}
\def\ri{\right)}
\def\les{\left[}
\def\ris{\right]}

\def\ric{\right\}}

\def\c#1{\cite{#1}}

\def\r#1{(\ref{#1})}

\def\epso{\epsilon_{\scriptscriptstyle 0}}
\def\lambdao{\lambda_{\scriptscriptstyle 0}}
\def\muo{\mu_{\scriptscriptstyle 0}}
\def\ko{k_{\scriptscriptstyle 0}}
\def\etao{\eta_{\scriptscriptstyle 0}}

\def\eps{\epsilon}
\def\epsa{\epsilon_a}
\def\epsb{\epsilon_b}
\def\epsc{\epsilon_c}

\def\epsmet{\eps_{met}}

\def\sp{{\mathbf s}}
\def\pinc{{\mathbf p}_+}
\def\pref{{\mathbf p}_-}

\def\Einc{{\mathbf E}_{inc}({\bf r})}
\def\Erefl{{\mathbf E}_{ref}({\bf r})}
\def\Etr{{\mathbf E}_{tr}({\bf r})}

\def\ctheta{\cos\theta}
\def\stheta{\sin\theta}

\def\nr{n_\ell}

\def\rs{r_s}
\def\rp{r_p}
\def\ts{t_s}
\def\tp{t_p}

\def\ux{\hat{\mathbf{u}}_x}
\def\uy{\hat{\mathbf{u}}_y}
\def\uz{\hat{\mathbf{u}}_z}

\begin{document}

\title{Sculptured--thin--film Plasmonic--Polaritonics}
\author{A. Lakhtakia,  J. A. Polo Jr., and M. A. Motyka
\thanks{Akhlesh Lakhtakia and Michael A. Motyka are with the Department of
Engineering Science and Mechanics, Pennsylvania State University,
University Park, PA 16802, USA, E-mail: akhlesh@psu.edu}
\thanks{John A. Polo Jr. is with the Department of Physics and Technology, Edinboro University of Pennsylvania,   235 Scotland Rd., Edinboro, PA  16444, USA, E-mail: polo@edinboro.edu
}}
\maketitle
\thispagestyle{empty}\pagestyle{empty}

\begin{abstract}The solution of a boundary--value problem formulated for the  Kretschmann configuration shows that the phase speed of a surface--plasmon--polariton
(SPP) wave
guided by the planar interface of a sufficiently thin metal film and a sculptured
thin film (STF)  depends on the vapor incidence angle used while fabricating the
STF by physical vapor deposition. Furthermore, it may be possible to engineer the phase
speed by periodically
varying the vapor incidence angle. The phase speed of the SPP wave
can be set by selecting  higher mean value and/or the  modulation  amplitude of the vapor
incidence angle.
\end{abstract}

\section{Introduction}
A resonance  phenomenon arises from the interaction of light with free electrons at a planar metal--dielectric interface \cite{ZSM,PSCE}. Under certain conditions, the energy carried by photons in the dielectric medium is transferred to collective excitations of free electrons in the metal.   Because the free electrons in the metal are coupled to the photons in the dielectric medium, the quantum
is called a \emph{surface plasmon polariton}, often shortened to \emph{surface
plasmon}. A classical understanding of the surface plasmon polariton (SPP) is in terms
of a electromagnetic surface wave that propagates along the interface and decays exponentially with distance normal to the interface.

Research on electromagnetic surface waves has a long history, dating back about
a hundred years.  Zenneck \cite{Zenneck} proposed in 1907 a mode of electromagnetic wave  propagation localized at the Earth--atmosphere interface.
The wave, since named the Zenneck wave, propagates parallel to the interface with an amplitude which decays exponentially with distance from the interface.  Credit is also given to  Sommerfeld  for a clean analysis of the phenomenon published in 1909 \cite{Sommerfeld}.  Basically the same kind of wave at optical frequencies, the SPP wave became the subject of intense investigation\cite{Raether,AgMills,Boardman} from the middle part of the 20th century.  It was soon realized that, because SPP waves are highly sensitive to conditions at the interface, they might be employed to detect various chemical species.  SPP--wave--detection techniques are currently used for a wide range of sensing applications especially in the detection of biomolecules \cite{LLS,HYG,KTGK,AZLe}.

Much of the SPP literature is restricted to the planar interface of a good conductor and  an isotropic dielectric medium, though there are noticeable exceptions wherein
the dielectric medium could be anisotropic \cite{WBBH,MBCTT} and be additionally
endowed with magnetic properties \cite{HBMBW,FB,YSRK}. The SPP wave is excited through one of several different types of couplers \cite{Knoll}.


\begin{figure}[!htb]
\centering \psfull
\epsfig{file=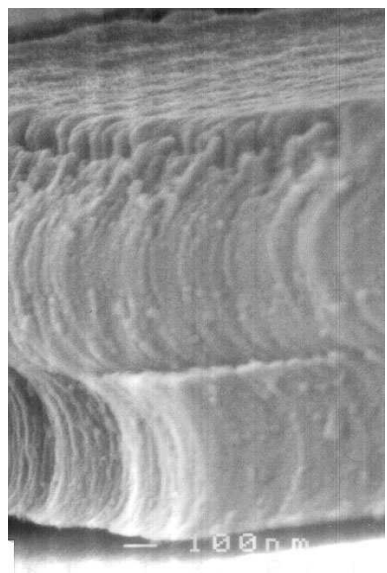,
width=5cm}
\caption{Scanning electron micrography of a sculptured nematic
thin film. The columnar morphology is essentially two--dimensional.
Courtesy: R.~Messier.
}
\label{figSNTFimage}
\end{figure}


In the Kretschmann \cite{KR68} configuration,
the bulk metal is in the form of a sufficiently thin film of uniform thickness, bounded by
dielectric mediums on both sides, one medium being optically denser
than the other.  A plane wave is
launched in the optically denser dielectric medium towards the metal film, in order to excite a surface--plasmon wave at the interface of the metal with the optically rarer dielectric medium. The plane wave must be $p$--polarized. The telltale sign
is a sharp peak in absorbance (i.e., a sharp trough in  reflectance
without a compensatory peak in transmittance) as the angle of incidence (with respect
to the thickness direction)
of the launched plane wave is changed. The absorbance peak occurs in the
vicinity of the critical angle (of incidence)   that would exist if the metal film were absent.

Generally, the
optically rarer medium is homogeneous normal to its planar interface
with the metal film, at least within the range of the
SPP field. In this paper, however, we take the optically rarer medium
to be
continuously
nonhomogeneous in the thickness direction. Specifically, this medium
is a sculptured thin film (STF).

\begin{figure}[!htb]
\centering \psfull
\epsfig{file=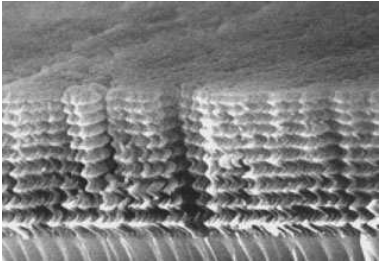,
width=5cm}
\caption{Scanning electron micrography of a chiral sculptured
thin film, which has a three--dimensional columnar morphology.
 Courtesy: R.~Messier.
}
\label{figchiSTFimage}
\end{figure}

STFs are nanostructured
materials with unidirectionally varying continuum properties that can be designed and realized in a
controllable manner using physical vapor deposition \cite{LMbook}. The ability to virtually
instantaneously change
the growth direction of their columnar morphology, through simple variations in
the direction of the incident vapor flux, leads to a wide variety  of $\sim100$--nm--diameter
columns
of two-- or three--dimensional shapes, as illustrated by the scanning electron
micrographs in Figs.~\ref{figSNTFimage} and \ref{figchiSTFimage}.
At visible and infrared wavelengths, a  STF is a unidirectionally nonhomogeneous
continuum with direction--dependent properties.


This paper is organized as follows. Section~\ref{theory} begins
with a description of the Kretschmann configuration to launch
a SPP wave at the planar interface of a metal and a STF. The
relative permittivity dyadic of the STF is described in that section,
along with the boundary--value problem to determine
planewave absorbance. Section~\ref{nrd} contains numerical
results to elucidate the effects of the nonhomogeneity of the
STF on the SPP wave.
An $\exp(-i\omega t)$ time--dependence is implicit, with $\omega$
denoting the angular frequency. The free--space wavenumber, the
free--space wavelength, and the intrinsic impedance of free space are denoted by $\ko=\omega\sqrt{\epso\muo}$,
$\lambdao=2\pi/\ko$, and
$\etao=\sqrt{\muo/\epso}$, respectively, with $\muo$ and $\epso$ being  the permeability and permittivity of
free space. Vectors are in boldface, dyadics underlined twice;
column vectors are in boldface and enclosed within square brackets, while
matrixes are underlined twice and similarly bracketed. Cartesian unit vectors are
identified as $\ux$, $\uy$ and $\uz$.

\section{Theory}\label{theory}
In conformance with the  Kretschmann configuration \cite{KR68,Simon} for launching SPP waves,
the half--space $z\leq 0$ is occupied by a homogeneous, isotropic, dielectric
material with the relative permittivity scalar $\eps_{\ell}$. Dissipation in this material
is considered to be negligible and its refractive index $n_\ell=\sqrt{\eps_\ell}$
is real--valued and positive. The laminar region
 $0 \leq z\leq L_{met}$ is occupied by a bulk metal with relative permittivity
 scalar $\epsmet$. The region
 $L_{met} \leq z \leq L_{met}+L_{stf}$
 is occupied by a STF.
  Without significant
 loss of generality in the present context, the half--space
 $z\geq L_{met}+L_{stf}$ is taken to be occupied by the same
 material as fills the half--space $z\leq 0$.

\subsection{Sculptured Thin Film}
 The
relative permittivity dyadic $\=\epsilon_{stf}(z)$
of the  STF   is
factorable as
\begin{eqnarray}
\nonumber
&&
\=\epsilon_{stf}(z) =  \=S_z(\zeta)\.\=S_y(\chi)\.\=\epsilon^{
ref}_{stf}
\.\=S_y^T(\chi)\.\=S_z^T(\zeta)
\, , \\
&&\qquad\qquad  L_{met}\leq z \leq L_{met}+L_{stf}\,,
\label{eps-STF}
\end{eqnarray}
wherein  the reference relative permittivity dyadic
\begin{equation}
\=\epsilon_{stf}^{ref}= \epsa  \, \uz\uz  + \epsb \, \ux\ux
+ \epsc \, \uy\uy\,
\end{equation}
captures the locally orthorhombic character of STFs,
the
dyadic function
\begin{eqnarray}
\nonumber
&&
\=S_z(\zeta)=\le \ux\ux + \uy\uy \ri \cos\zeta\\
&&\qquad
+\le \uy\ux -
\ux\uy \ri \sin\zeta+\uz\uz \,
\end{eqnarray}
denotes rotation about the $z$ axis, the dyadic function
\begin{eqnarray}
\nonumber
&&
\=S_y(\chi)=\le \ux\ux + \uz\uz \ri \cos{\chi}
\\
&&\qquad+\le \uz\ux -
\ux\uz \ri \sin{\chi}+\uy\uy \,
\end{eqnarray}
involves the angle   $\chi\in[0,\pi/2]$, and
the superscript $^T$ denotes the transpose.
The quantities $\zeta$, $\chi$, $\epsilon_a$, $\epsilon_b$,
and $\epsilon_c$ can all be functions of $z$.

\begin{figure}[!htb]
\centering \psfull
\epsfig{file=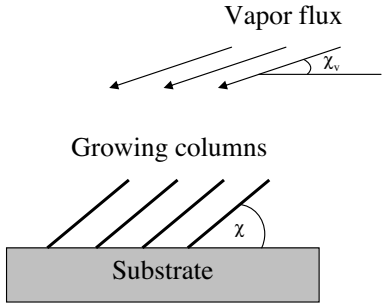,
width=5cm}
\caption{Schematic of the growth of a columnar thin film. The vapor flux is directed at an angle $\chi_v$,
whereas columns grow at an angle   $\chi\geq\chi_v$.
}
\label{figCTFgrowth}
\end{figure}

Although STFs have been made by evaporating a wide variety of materials
\cite{LMbook,LDHX}, the
constitutive parameters  of STFs have not been extensively measured.
However, the constitutive
parameters of certain columnar thin films (CTFs) are known.
Both CTFs and STFs are fabricated by
physical vapor deposition. The basic procedure to deposit CTFs has been known for more than a century. At low enough temperature and pressure, a solid material confined in a boat evaporates
towards a stationary substrate.  The vapor flux is collimated into a well--defined beam, and its average direction  is quantified by the angle
$\chi_v\in(0,\pi/2]$ with respect to the substrate plane,
as illustrated
in Fig.~\ref{figCTFgrowth}.   Provided the adatom mobility is low, the
resulting film turns out to be an assembly of parallel and nominally identical columns. The
columns have elliptical cross--sections and are tilted at an angle $\chi \geq \chi_v$ with respect to
the substrate plane.  The parameters  $\chi$, $\epsilon_a$, $\epsilon_b$,
and $\epsilon_c$ have to be functions of
$\chi_v$, at the very least because the nanoscale porosity of a CTF depends on the direction of the
vapor flux.


A series of optical
characterization experiments on certain  CTFs   were carried out
some years ago \c{HWH,HWC}. After ignoring the effects of dispersion
and dissipation, at least in some narrow range of frequencies, the results can be
put in the following form for our present purposes:
\begin{equation}
\label{Hodg1}
\left. \begin{array}{ll}
\eps_a  = \le n_{a0} + n_{a1}\,v + n_{a2}\,v^2\ri^2\\[5pt]
\eps_b = \le n_{b0} + n_{b1}\,v + n_{b2}\,v^2\ri^2\\[5pt]
\eps_c  = \le n_{c0} + n_{c1}\,v + n_{c2}\,v^2\ri^2\\[5pt]
\chi = \tan^{-1}\left(m\,\tan\chi_v\right)
\end{array}\ric\,.
\end{equation}
Here, $v =\chi_v/(\pi/2)$
is the vapor incidence angle expressed as a fraction of a right
angle. The quantities $m$ and $n_{a0}$, etc., in Eq.~\ref{Hodg1}
depend on the
evaporant material as well as the deposition conditions.

When the substrate is rotated about either the $y$ or the $z$ axes,    parallel
columns of specific shape grow instead of
straight columns, and a STF is deposited instead of a CTF. Although
the substrate is nonstationary, the functional relationships
connecting  $\chi$, $\epsilon_a$, $\epsilon_b$,
and $\epsilon_c$ to $\chi_v$ for CTFs would
substantially apply
for STFs, provided the substrate rotation is sufficiently slow.
Thus, we need only to specify the $z$--dependences
of $\zeta$ and $\chi_v$.

For our present purposes, we chose
\begin{equation}
\left.\begin{array}{l}
\zeta(z) =h\, \frac{\pi}{\Omega}\left(z-L_{met}\right)\\[5pt]
\chi_v(z)=\tilde{\chi}_v +\delta_v\,\sin\left[\frac{\pi}{\Omega}\left(z-L_{met}\right)
\right]
\end{array}\right\}\,.
\end{equation}
Here, $\Omega$ is a characteristic length along the $z$ axis, whereas
the angles $\tilde{\chi}_v\in\left(0,\pi/2\right]$ and $\delta_v\in\left[0,
\tilde{\chi}_v\right]$. The structural--handedness parameter $h=1$
for right--handedness, $h=-1$ for left--handedness, and
$h=0$ for no handedness. For theoretical investigations,
we decided to focus on the
following three types of STFs:
\begin{itemize}
\item[(i)] columnar thin films ($h=0$, $\delta_v=0$);
\item[(ii)] sculptured nematic thin films with periodically
varying tilt angle ($h=0$, $\delta_v>0$); and
\item[(iii)] chiral sculptured thin films
($h=\pm 1$, $\delta_v=0$).
\end{itemize}

\subsection{Boundary--Value Problem}
Suppose that a $p$--polarized plane
wave, propagating in the half--space
$z \leq 0$ at an angle $\theta\in[0,\pi/2)$ to the $z$ axis  in the $xz$ plane, is incident on the metal--coated STF in the Kretschmann configuration.
The electric field phasor associated
with the incident plane wave is
\begin{equation}
\Einc=  \pinc \, e^{ i\kappa x
} \,e^{i\ko\nr z\ctheta}
\, , \qquad z \leq 0
\, .
\end{equation}
The reflected electric field phasor is expressed as
\begin{eqnarray}
\nonumber
&&
\Erefl= (\rs\,\sp +\rp \,\pref) \, e^{ i\kappa x
}
\\&& \qquad \times\,e^{-i\ko\nr z\ctheta}\,,\qquad z \leq 0
\, ,
\end{eqnarray}
and the transmitted electric field phasor  as
\begin{eqnarray}
\nonumber
&&
\Etr= (\ts\,\sp +\tp\, \pinc) \, e^{ i\kappa x
} \\
&&\nonumber\qquad\times\,
e^{i\ko\nr (z-L_{met}-L_{stf})\ctheta}
\, ,\\
&&\qquad \qquad z \geq L_{met}+L_{stf}
\, .
\end{eqnarray}
Here,
\begin{equation}
\left.\begin{array}{l}
\kappa =
\ko\nr\stheta\\[5pt]
\sp=\uy
\\[5pt]
{\mathbf p}_\pm=\mp  \ux   \ctheta  + \uz \stheta
\end{array}\right\}
\, ,
\end{equation}
where $\omega/\kappa$ is the phase speed parallel to the interfacial
plane $z=L_{met}$ of interest, and the unit vectors $\#s$ and $\#p_{\pm}$ denote
the $s$-- and the $p$--polarization states of the electric field phasors.

The reflection amplitudes $\rs$ and $\rp$, as well as the transmission
amplitudes $\ts$ and $\tp$, have to be determined by the solution of
a boundary--value problem. The required procedure  is standard
\cite{LMbook}.
 It suffices to state here that the following set
of four algebraic equations emerges (in matrix notation):
\begin{eqnarray}
\nonumber
&&\left[\begin{array}{l} t_s\\t_p\\ 0 \\0\end{array}\right]=
[\=K]^{-1}\cdot[\=M_{stf}]
\\
&&\qquad
\cdot
\exp\left({i[\=P_{met}]L_{met}}\right)\cdot[\=K]\cdot
\left[\begin{array}{l} 0\\ 1\\ r_s \\r_p\end{array}\right]\,.
\label{finaleq}
\end{eqnarray}
The procedure to obtain the 4$\times$4 matrix
$[\=M_{stf}]$ can be gleaned from two predecessor papers \cite{PLsntf,LakhSPR}.
The remaining two
4$\times$4 matrixes in  Eq. \ref{finaleq} are as follows:
\begin{equation}
[\=K] =
\left[ \begin{array}{cccc}
0 & - \cos\theta  & 0 &  \cos\theta \\
1 & 0  & 1 & 0 \\
-\left(\frac{\nr}{\etao}\right) \cos\theta & 0
&\left(\frac{\nr}{\etao}\right) \cos\theta &
0 \\
0 &
-\, \frac{\nr}{\etao}  & 0 & -\, \frac{\nr}{\etao}
\end{array}\right]\,,
\end{equation}
\begin{eqnarray}
\nonumber
[\=P_{met}]&=&\les\begin{array}{cccc}
0 & 0 & 0 & \omega\muo \\[4pt]
0 & 0 & -\omega\muo & 0 \\[4pt]
0 & -\omega\epso\epsmet & 0 &0\\[4pt]
\omega\epso\epsmet &0 & 0 & 0
\end{array}\ris\\[5pt]
 &&
+
\les\begin{array}{cccc}
0 & 0 & 0 & -\frac{\kappa^2}{\omega\epso\epsmet}
\\[4pt]
0 & 0 & 0  &0 \\[4pt]
0 & \frac{\kappa^2}{\omega\muo}& 0 & 0\\[4pt]
0 & 0 & 0 & 0
\end{array}\ris\,.
\label{Pmetal}
\end{eqnarray}

Equation \ref{finaleq} can be solved for $\rs$, $\rp$, $\ts$, and $\tp$
using standard algebraic techniques. The   quantity of interest
for establishing the existence of the SPP wave is the absorbance
\begin{equation}
A_p=1-\left(\vert r_s\vert^2+\vert r_p\vert^2+\vert t_s\vert^2+\vert t_p\vert^2\right)\,
\end{equation}
as a function of $\theta$.

\section{Numerical Results and Discussion}\label{nrd}
Calculations of $A_p$ against $\theta$ are reported in this paper  for the
following parameters of the STF: $\Omega=200$~nm, $L_{stf}=4\Omega$, and
$\tilde{\chi}_v=30^\circ$. Calculations have shown
that the selection of higher values of $L_{stf}$
does not impact the SPP wave significantly, whereas the effect of the
nanostructured periodicity of SNTFs and chiral STFs is not appreciable for lower values of $L_{stf}$.
The free--space wavelength
$\lambdao=633$~nm is the same at which the parameters in Eqs.~\ref{Hodg1} were measured.
The STF was taken to be made of titanium
oxide: $n_{a0}=1.0443$, $n_{a1}=2.7394$, $n_{a2}=-1.3697$,
$n_{b0}=1.6765$, $n_{b1}=1.5649$, $n_{b2}=-0.7825$,
$n_{c0}=1.3586$, $n_{c1}= 2.1109$, $n_{c2}=-1.0554$,
and $m=2.8818$ \cite{HWH}. The metal was chosen to be aluminum: $\epsmet=-56+i21$ \cite{Mansuripur} and
$L_{met}=15$~nm. The metal film is thus thin enough 
that it allows sufficient penetration of the evanescent wave to excite the SPP at the metal--STF interface; at the same time, the metal film is thus thick
enough to prevent tunneling of photons across it from the medium
of incidence and reflection to the STF. The two half--spaces were taken to be filled with zinc selenide
($\nr=2.58$), which is optically denser than all three types of STFs
considered here.

\begin{figure}[!htb]
\centering \psfull
\epsfig{file=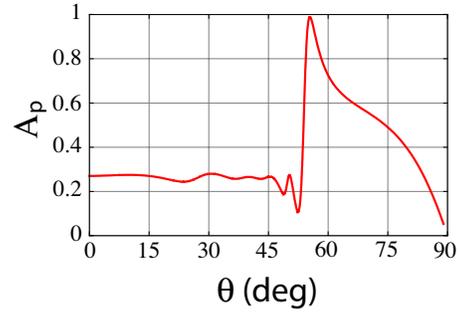,
width=6cm}
\caption{Absorbance $A_p$  as a function of   $\theta$ when
the STF is a columnar thin film ($h=0$ and $\delta_v=0$);
see the text for other parameters. The SPP wave is excited at
$\theta=55.41^\circ$.
}
\label{figCTF}
\end{figure}

Let us begin with the case of the metal--backed CTF: $h=0$ and $\delta_v=0$. Figure~\ref{figCTF} shows $A_p$ as a function of   $\theta$. The
SPP wave is excited at $\theta=55.41^\circ$, with the absorbance
in excess of $0.988$ denoting a very efficient conversion of the
energy of the incident plane wave into the SPP wave.


Let $\theta_{spp}$ denote the value of $\theta$ at which the SPP wave
is excited. A study of $\theta_{spp}$ versus $\chi_v$
reveals that $\theta_{spp}$ for a CTF fabricated with a specific evaporant material
increases as $\chi_v$ increases \cite{LPajp}. Therefore, the wavenumber of the SPP wave, given by
\begin{equation}
\kappa_{spp}=\ko n_\ell\,\sin\theta_{spp}
\,,
\end{equation}
 is a monotonically increasing function of $\chi_v$, which means that the phase speed
\begin{equation}
v_{spp}=\omega/\kappa_{spp}
\end{equation}
of the SPP wave
is a monotonically decreasing function of $\chi_v$.


\begin{figure}[!htb]
\centering \psfull
\epsfig{file=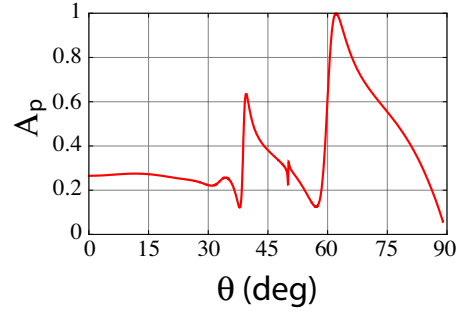,
width=6cm}
\caption{Absorbance $A_p$  as a function of   $\theta$ when
the STF is a sculptured nematic thin film ($h=0$ and $\delta_v=20^\circ$);
see the text for other parameters.  The SPP wave is excited at
$\theta=62.13^\circ$. The sharp changes for two values
of $\theta < 50^\circ$ do not indicate the excitation of SPP
waves.
}
\label{figSNTF}
\end{figure}


The effect of nonhomogeneity in the dielectric medium on the SPP wave
becomes evident when we consider the metal--backed sculptured nematic
thin film:
$h=0$ and $\delta_v>0$. Figure~\ref{figSNTF} shows $A_p$ as a function of   $\theta$ when $\delta_v=20^\circ$.
The
SPP wave is excited at $\theta=62.31^\circ$, with the absorbance
in excess of $0.998$. A comparison of Figs.~\ref{figCTF}
and \ref{figSNTF} suggests that the phase speed of the SPP wave decreases
when the vapor incidence angle $\chi_v$ is periodically modulated with a significant
amplitude, but the high
efficiency of conversion of the
energy of the incident plane wave into the SPP wave is not affected thereby.


\begin{figure}[!htb]
\centering \psfull
\epsfig{file=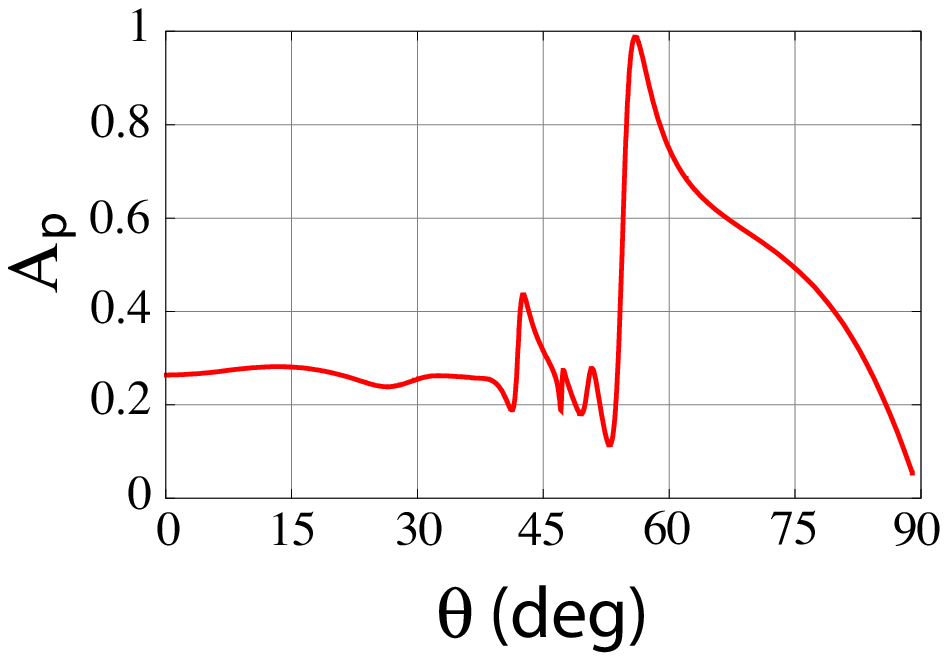,
width=6cm}
\caption{Absorbance $A_p$  as a function of   $\theta$ when
the STF is a    chiral sculptured thin film ($h=\pm1$ and $\delta_v=0$);
see the text for other parameters.  The SPP wave is excited at
$\theta=55.97^\circ$.
}
\label{figchiSTF}
\end{figure}


Nonhomogeneity is also introduced in the dielectric medium when the
 vapor deposition angle is held fixed during fabrication but the substrate
 is rotated about the $z$ axis. Figure~\ref{figchiSTF} shows $A_p$ as a function of   $\theta$ when $\delta_v=0$ and $h=\pm1$; i.e., it is drawn for
 a metal--backed chiral STF. Now, $\theta_{spp}=55.97^\circ$, which is only
 marginally higher than in Fig.~\ref{figCTF}.

 \section{Concluding Remarks}
We conclude that the solution of a boundary--value problem formulated for the
Kretschmann configuration shows that the phase speed of a SPP wave
guided by the planar interface of a sufficiently thin metal film and a sculptured
thin film  depends on the vapor incidence angle used while fabricating the
STF by physical vapor deposition. Therefore, it may be possible to engineer the phase
speed quite simply by selecting an appropriate value of the vapor incidence angle (in
addition to the metal and the evaporant species). Furthermore, by periodically
varying the vapor incidence angle, the phase speed of the SPP wave
can be reduced. Adequate selection  of the phase speed should be important
for controlled data flow in plasmonic circuits.

The high degree of porosity \cite{LMbook} of  STFs may provide certain advantages in the application of SPP waves in detectors.  A properly chosen STF could prevent particulates from reaching its interface with the metal film while allowing molecular species through for detection.  The detection of biomolecules typically relies on recognition molecules, which are bound to the interface, binding with the analyte
molecules.  The porosity and surface roughness of the STF may offer some advantage for adherence of the recognition molecule.  STFs can now be patterned using standard photolithographic techniques \cite{Horn04}.  Partitioning of the STF into many sectors could permit the use of multiple species of  recognition molecules on a single chip for  the detection of many different types of analyte molecules.  Recently, the activity of living cells has been monitored using SPP detection techniques \cite{Ziblat}.  Cells have been shown to grow well on the rough STF surface \cite{LDHX,Demirel07}, which thus may offer an advantage.  Finally, titanium oxide is a very well--known photocatalyst.  Although the photocatalytic properties of titanium oxide STFs \cite{Suzuki01,Wein95} have been investigated, the combination of SPP detection and this catalyst may allow for other useful applications to emerge.

\end{document}